\documentstyle[epsf]{mn}

\def\gsim{\;\lower.6ex\hbox{$\sim$}\kern-7.75pt\raise.65ex\hbox{$>$}\;}
\def\lsim{\;\lower.6ex\hbox{$\sim$}\kern-7.75pt\raise.65ex\hbox{$<$}\;}
\def\i814{I$_{\rm 814}$}
\def\v555{V$_{\rm 555}$}

\title[RR Lyraes in M 3]{ RR Lyrae variables in the globular cluster M3 
(NGC5272). I. BVI CCD photometry }

\author[Carretta et al.]
{E. Carretta$^{1,2}$,
C. Cacciari$^{2,3}$,
F.R. Ferraro$^2$,
F. Fusi Pecci$^{2,4}$,
G. Tessicini$^2$\\
$^1$ Osservatorio Astronomico di Padova, Vicolo dell'Osservatorio 5, 
I-35122 Padova, Italy; e-mail: carretta@astrpd.pd.astro.it\\
$^2$ Osservatorio Astronomico, via Zamboni 33, I-40126 Bologna, 
Italy; e-mail: carretta, cacciari, ferraro, flavio,
tessicini@astbo3.bo.astro.it\\
$^3$ ESA Space Science Department, Space Telescope Science Institute, 
3700 San Martin Dr., Baltimore, MD21218, USA\\
$^4$ Stazione Astronomica di Cagliari, 09012 Capoterra, Italy
 }

\date{}

\begin{document}
\maketitle

\begin{abstract}
New BVI CCD photometry is presented for 60 RR Lyrae variables in the 
globular cluster M3. Light curves have been constructed and ephemerides 
have been (re)-derived for all of them. Four stars (i.e. V29, V136, V155 and 
V209), although recognized as variables, had no previous period 
determinations. Also, the period derived for V129 is significantly different 
from the one published by Sawyer-Hogg (1973). Light curve parameters, 
i.e. mean magnitudes, amplitudes and rise-times, have been derived. 

The discussion of these results in the framework of the stellar evolution 
and pulsation theories will be presented in a forthcoming paper.

\end{abstract}


%
%-----------------------------------------------------------------------
%-----------------------------------------------------------------------
%
\section{Introduction}
Multi-colour observations of RR Lyrae variables are extremely important 
to obtain information on their physical parameters, such as 
temperature, mass, and luminosity, which are essential to the understanding 
of the different characteristics of these stars in different clusters. 
One of the most significant differences is related to the separation 
of globular 
clusters into two groups according to the mean period of the RRab variables, 
which was found by Oosterhoff (1939, 1944), namely $< P_{ab} >$=0.55 days 
for group I (OoI) and $< P_{ab} >$=0.65 days for group II (OoII). 
This ``dichotomy'', later found to be related to the cluster metallicity 
(Arp 1955), has been the basis for many subsequent studies which have led 
several authors, {\it in primis} 
Sandage (1981, 1990, 1993 and references therein) to define a 
luminosity-metallicity 
relation among RR Lyrae stars as a result of this observed ``period-shift''. 
The fiducial cluster in the study of this period-shift effect has 
traditionally been M3 for the OoI group. The data for the M3 RR Lyraes, 
however, were still essentially the UBV 
photographic photometry by Roberts and Sandage (1955), Baker and Baker (1956) 
and Sandage (1959), therefore a program was started 
in order to obtain more accurate CCD photometry for 
the study of the characteristics of both the Colour-Magnitude Diagram and 
the RR Lyrae variables in this cluster. 

M3 (NGC5272; RA=13 42 11, DEC=+28 22 32 (2000); l=42$\deg$ b=79$\deg$) 
is among the most important and prominent clusters in the 
northern hemisphere. It is located at $\sim$ 11.6 kpc from the 
Galactic Center and 9.5 kpc above the plane, and is at 9.7 kpc from the 
Sun (Harris 1996). 
It contains the largest number of variable stars known within a single 
cluster (260 are quoted by Suntzeff, Kinman \& Kraft 1991) with a rather 
high specific frequency of RR Lyraes (i.e. normalized to total mass). 
The frequency of double-mode pulsators, however, is quite low: only 
three RRd are known (Nemec \& Clement 1989, hereafter NC89, and Kaluzny 
et al. 1997, hereafter K97), two of which (V68 and V87) have been known 
to be RRd stars for some time, and the third one (V79) has been identified 
as an RRd star only recently because it appears to have switched modes some 
time between 1962 and 1996 (Clement et al. 1997). 
Its metallicity, for a long time considered to be [Fe/H] $\sim$ -1.6, 
has been recently revised using high resolution spectroscopic 
data and appears to be somewhat higher, i.e. $\sim$ -1.4 
(Carretta \& Gratton 1997). 

M3 has been the target of a very large number of studies and analyses. 
For what concerns 
the photometry of RR Lyrae variables, we mainly refer to Roberts \& Sandage 
(1955), Baker \& Baker (1956) and Sandage (1959), which constituted 
the largest body of photometric data and was used as a basis 
for all subsequent studies of the global properties of RR Lyraes. 
Several more references 
on individual stars are listed in Sawyer-Hogg (1973 - hereafter SH73)  
and in the updated compilation of the SH73 Catalogue by Clement 
(1996). The most recent CCD observations of RR Lyraes in M3 are 
from K97 and from M. Corwin (private communication). 

In this paper we present a new photometric study of 60 RR Lyrae variable 
stars in M3, based on BVI CCD observations taken over a period of 5 
years and a new independent absolute photometric calibration (described in 
detail by Ferraro et al. 1997, hereafter F97). 
Here we will present the observations and an initial analysis of the data. 
A more detailed analysis of the RR Lyrae properties is deferred to 
a separate paper (Cacciari et al. 1998), whereas the new photometry 
of the extended sample of non-variable stars ($\sim$ 19,000 stars) 
and related CMD have been presented and discussed by Buonanno et al. 
(1994: hereinafter B94) and F97. 

\section{Observations and Data Reduction}

The observations were obtained at two sites, Loiano (Bologna Observatory, 
Italy) and the German-Spanish Astronomical Center in Calar Alto (Spain). 

\begin{figure}
\vspace{12cm}
\caption{ The M3 SE field observed from Loiano. Variables are identified with 
their SH73 name.}
\includegraphics{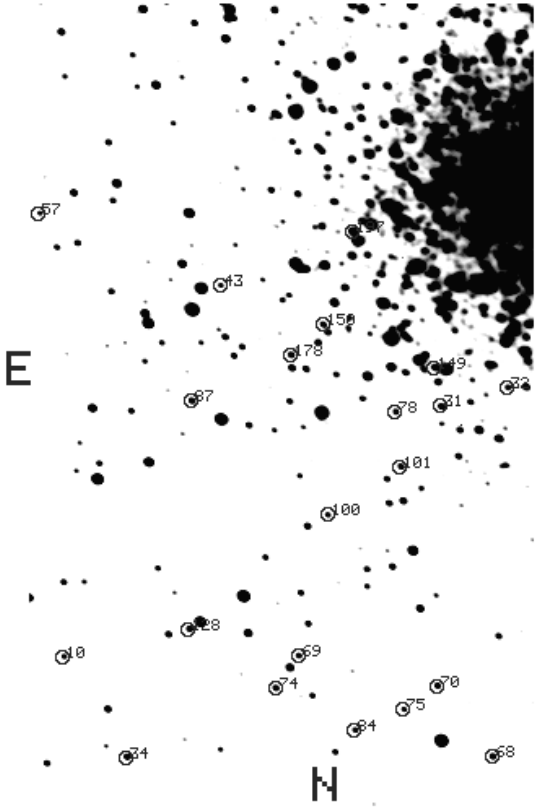}
\label{fig-f1}
\end{figure}

\begin{figure}
\vspace{12cm}
\caption{ As in Figure 1, but for the M3 NW field observed from Loiano.}
\includegraphics{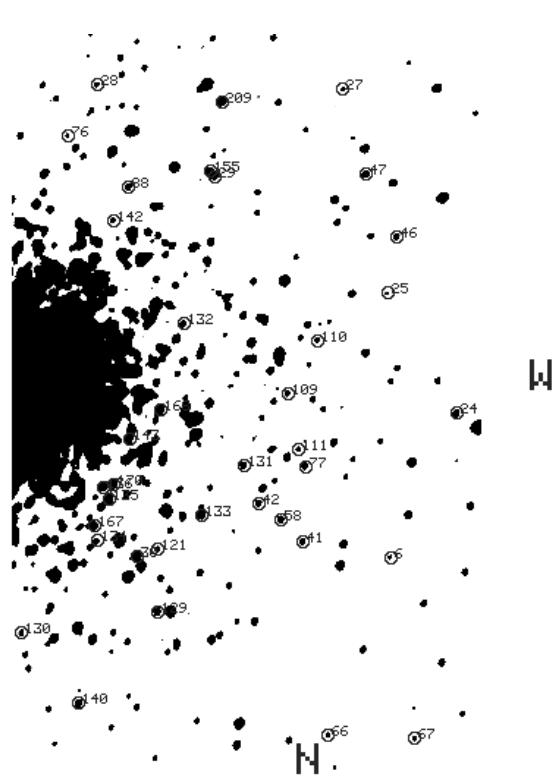}
\label{fig-f2}
\end{figure}

The Loiano observations were taken in: March, 3-5, 7-9, and 11 1990; 
February, 7 1992; April, 10-11 1992; and May, 3, 7-9, and 11 1992 
with the 152cm F/8 Ritchey-Chretien telescope and 
an RCA CCD 320x512 pixels, scale =0.5arcsec/pixel, FOV=2'40"x4'16". 
Because of the limited size of the field of view, two contiguous fields 
were observed (see Figure~\ref{fig-f1} and Figure~\ref{fig-f2}) 
to cover the SE and NW halves of 
the cluster. 
Even so, the most external variables could not be observed.

The standard Johnson BVI filters were used, and 65 and 69 frames in each 
colour were taken on the NW and SE field respectively. 
The exposure times depended on conditions, and were typically
 $\sim$ 5 min (V), 10-15 min (B) and 4 min (I) for an average 
seeing of about 1.5-2 arcsec. In a few cases, however, the seeing 
was significantly worse and the data have been discarded from the subsequent 
analysis. 

A total of 38 variables in the NW field and 22 variables in the SE field 
have been measured based on previously identified positions (SH73), 
and for several of them 
improved or new periods have been determined (see Sect. 3.). 
No independent search for variability has been performed. 

The Calar Alto observations are only 5 data points spread over three 
almost contiguous nights (March, 26 1995 and April, 1,2 1995), and were 
taken mainly for calibration purposes. 
However they provide a good extension of the time baseline and have proven 
quite helpful in the period analysis. 
The observations were taken with the 1.23 m telescope, using a thinned 
1024x1024 Tektronix chip (24 $\mu$ pixel, 0.50 arcsec/pixel), with 
Ar-coatings. The I filter is in the Kron-Cousins system, centered at 
8020 \AA. 
For an average seeing of 1.6 arcsec the typical exposure times were 
$\sim$ 6-7 min (V), 20 min (B) and 5-6 min (I). 

The data reduction and absolute photometric calibration of the RR Lyrae data 
is the same that was obtained and used for the construction of the new 
Colour-Magnitude diagram of M3 (F97). The absolute calibration in particular, 
which was the basic item for a new and independent analysis of the M3 
characteristics, turned out to be a rather complicated and difficult 
problem to solve with the desired accuracy. Since it has been discussed 
in great detail by F97, along with the comparison with all previous 
relevant photometries, we refer the interested reader 
to this paper; we shall quote and use their results when relevant. 

The photometry of the RR Lyrae variables was carried out relatively to 
the local standard sequence that had been defined in each field, so 
as to cover the colour and magnitude range relevant for HB stars. 
The RR Lyraes we have measured are identified in Figure 1. 
The local standard stars are listed in Table~\ref{tab-locstan}, where 
we give the photometric values with the new calibration of F97. 
Star identifications are from B94.

\begin{table*}
\caption{ Local standard stars. The identifications are taken from 
Buonanno et al. (1994)}
\begin{center}
\begin{tabular}{lccclccclccc}
\hline\hline
\multicolumn{4}{c}{Field West}  &\multicolumn{4}{c}{Field East}  
&\multicolumn{4}{c}{Field East} \\
Star &   B    &   V    &   I    & Star &   B    &   V    &   I    & Star &   B
   &   V    &   I   \\
\hline
284  & 16.681 & 15.877 & 14.975 & 201  & 15.922 & 15.798 & 15.704 & 472  &
15.961 & 15.745 & 15.538\\
293  & 16.332 & 15.495 & 14.568 & 202  & 16.163 & 15.633 & 14.911 & 477  &
15.556 & 14.763 & 13.838\\
304  & 15.196 & 14.235 & 13.177 & 210  & 16.905 & 16.144 & 15.208 & 521  &
17.589 & 16.870 & 16.011\\
305a & 15.801 & 15.581 & 15.413 & 216  & 17.153 & 16.429 & 15.495 & 522b &
17.311 & 16.595 & 15.759\\
331  & 16.254 & 15.453 & 14.551 & 230  & 16.170 & 15.334 & 14.325 & 582  &
15.824 & 15.639 & 15.519\\
338  & 16.190 & 15.381 & 14.453 & 233  & 16.605 & 15.826 & 14.894 & 590  &
17.315 & 16.574 & 15.711\\
353  & 16.187 & 15.745 & 15.233 & 233a & 17.682 & 16.983 & 16.075 & 600  &
15.666 & 14.756 & 13.755\\
354  & 16.002 & 15.395 & 14.672 & 250  & 17.426 & 16.712 & 15.843 & 619  &
17.414 & 16.712 & 15.875\\
346  & 17.229 & 16.591 & 15.834 & 252  & 17.492 & 16.844 & 15.953 & 622  &
16.554 & 15.804 & 14.920\\
348  & 16.641 & 15.867 & 14.956 & 273  & 16.190 & 15.635 & 14.963 & 624  &
16.382 & 15.608 & 14.668\\
394  & 15.627 & 14.941 & 14.123 & 281  & 15.750 & 14.875 & 13.855 & 655  &
16.054 & 15.228 & 14.273\\
409  & 15.838 & 14.986 & 14.027 & 282a & 16.193 & 15.626 & 14.928 & 662a &
17.687 & 16.967 & 16.146\\
431  & 15.589 & 14.717 & 13.764 & 283  & 16.061 & 15.478 & 14.744 & 688  &
16.029 & 15.566 & 15.015\\
444  & 15.715 & 15.506 & 15.327 & 300  & 17.585 & 16.874 & 16.018 & 699  &
16.788 & 16.051 & 15.228\\
460  & 17.153 & 16.420 & 15.597 & 302  & 16.915 & 16.202 & 15.338 & 726  &
15.541 & 14.647 & 13.637\\
486a & 15.327 & 14.481 & 13.542 & 325  & 17.166 & 16.473 & 15.620 & 727  &
16.135 & 16.101 & 16.187\\
520  & 15.421 & 14.602 & 13.692 & 336  & 16.903 & 16.169 & 15.259 & 730  &
16.902 & 16.182 & 15.317\\
559  & 16.285 & 15.472 & 14.596 & 337  & 16.445 & 15.700 & 14.879 & 733  &
16.062 & 16.022 & 16.090\\
581  & 16.207 & 15.681 & 15.134 & 347  & 16.667 & 15.897 & 14.962 & 750  &
16.762 & 15.970 & 15.095\\
594  & 17.098 & 16.397 & 15.643 & 355  & 15.209 & 14.250 & 13.183 & 769  &
16.044 & 15.209 & 14.247\\
598  & 16.050 & 15.990 & 16.094 & 356  & 16.297 & 15.812 & 15.196 & 772  &
16.054 & 15.499 & 14.864\\
598a & 16.860 & 16.127 & 15.310 & 362  & 17.524 & 16.847 & 16.058 &&&&\\    
641  & 16.057 & 15.199 & 14.298 & 384  & 16.058 & 15.393 & 14.605 &&&&\\
636a & 15.864 & 15.693 & 15.669 & 422  & 16.793 & 16.025 & 15.155 &&&&\\
653  & 15.959 & 15.562 & 15.140 & 434  & 16.258 & 16.264 & 16.351 &&&&\\
670  & 15.386 & 14.641 & 13.836 & 437  & 15.183 & 14.263 & 13.224 &&&&\\
708  & 15.704 & 14.815 & 13.867 & 438  & 15.870 & 15.738 & 15.576 &&&&\\
770  & 15.831 & 15.665 & 15.699 & 448a & 16.416 & 15.587 & 14.647 &&&&\\
776  & 16.025 & 15.441 & 14.818 & 449  & 17.103 & 16.370 & 15.474 &&&&\\
\hline
\end{tabular}
\label{tab-locstan}
\end{center}
\end{table*}

The {\it internal} error of each measurement for the RR Lyrae stars 
depends on the seeing and crowding conditions, and is on average 
about 0.02-0.03 mag. As far as {\it systematic} errors are concerned, 
we recall the conclusion by F97 (see their Sect. 2.5) that 
`` we still cannot exclude errors as high as 0.05 mag in the absolute 
values, both in magnitude and in colour (particularly at the blue and 
red extremes)''. 

A comparison with the photometry of K97 for the non-variable stars in the 
magnitude interval V=15.0 to 16.5 shows that our V data are comparable 
to K97 within the errors. In B-V there is a colour equation 
$ (B-V)_{F97} = (B-V)_{K97} - 0.105(B-V)_{K97} +0.094$, 
which implies a difference $\Delta B \sim$ 0.06 mag (our data minus K97) 
at the colour interval of the RR Lyraes. 
This is consistent with the comparison of the average V 
magnitudes for the 9 variable stars we have in common, which shows that 
the average V magnitudes are the same within 
0.05 mag in the worst case. Considering that i) K97 do not exclude the 
possibility of a systematic difference in their own photometry by up to 
0.02 mag between the north and south fields they have monitored, ii) some 
stars have intrinsic variations in the shape of their light curves, and 
iii) the average V magnitudes were calculated in different ways, we can 
conclude that the photometric properties of our data-sets in the magnitude 
and colour ranges relevant for the RR Lyrae variables are the same, 
within the observational errors, for what concerns the V filter. The B data, 
however, may be affected by some systematic differences. 

The magnitudes we have derived are presented 
in Table~\ref{tab-varmag}, along with the Heliocentric Julian Day 
corresponding to the beginning of the exposure. 
The variables are identified as in SH73. 
In order to save space, we present in Table~\ref{tab-varmag} only the 
B data for a few variable stars observed in the West field, as an example. 
Similar data for all of the variables observed, both in the East and West 
fields, as well as in the V and I filters, are available from the first 
author upon request.

\begin{table*}
\caption{ Example of the B data for some of the variable stars in the
West field. The Heliocentric Julian Days correspond to the beginning of 
the exposure.} 
\begin{center}
\begin{tabular}{lcccccccccc}
\hline\hline
%\multicolumn{10}{c}{ B photometry. Field west variables} \\
      JHD    &  6   &  24  &   25 & 27   & 28   & 29   & 30   & 41   & 42    \\
\hline         
 47954.44854 &16.326&15.726&16.476&16.256&15.126&16.226&15.556&16.116&15.596\\
 47954.46452 &16.366&15.816&16.556&16.466&15.206&16.196&15.656&16.126&15.666\\
 47954.48257 &16.426&15.836&16.516&16.276&15.226&16.176&15.736&16.136&15.746\\
 47954.49924 &16.446&15.876&16.466&16.296&15.316&16.146&15.806&16.136&15.816\\
 47954.51591 &16.466&15.956&16.496&16.336&15.476&16.056&15.866&16.146&15.876\\
 47954.53327 &16.456&15.986&16.436&16.366&15.626&15.766&15.946&16.166&15.966\\
 47954.55202 &16.436&16.046&16.526&16.446&15.806&15.646&15.996&16.176&16.016\\
 47954.57077 &16.416&16.076&16.466&16.356&15.906&15.486&16.066&16.176&16.086\\
 47954.58882 &16.406&16.066&16.456&16.346&15.886&15.466&16.056&16.166&16.076\\
 47954.60757 &16.396&16.136&16.586&16.496&16.136&15.466&16.196&16.116&16.156\\
 47954.62771 &16.396&16.186&16.576&16.546&16.206&15.526&      &16.116&16.206\\
 47954.64924 &16.466&16.216&15.946&16.576&16.286&15.676&      &16.086&16.196\\
 47954.67007 &16.506&16.236&15.026&16.586&16.276&15.826&      &15.986&16.226\\
 47955.39716 &16.006&16.206&16.426&15.676&15.146&16.196&15.076&16.016&16.366\\
 47955.42146 &16.106&16.216&16.496&      &15.266&16.276&15.096&16.056&16.406\\
 47955.44507 &16.206&16.196&16.486&15.786&15.436&16.176&15.316&16.116&16.336\\
 47955.46452 &16.256&16.256&16.536&15.976&15.586&16.136&15.446&16.126&15.896\\
 47955.48327 &16.306&16.306&16.496&16.056&15.766&15.866&15.576&16.136&15.206\\
 47955.50271 &16.376&16.316&16.446&16.086&15.856&15.596&15.666&16.126&14.796\\
 47955.52146 &16.446&16.346&16.446&16.116&15.956&15.426&15.766&16.146&14.866\\
 47955.54021 &16.436&16.346&16.416&16.176&      &15.396&15.846&16.126&14.986\\
 47955.58327 &16.426&16.086&16.516&16.286&16.186&15.556&16.006&16.116&15.266\\
 47955.60271 &16.416&15.736&16.106&16.276&16.186&15.656&16.086&16.076&15.376\\
 47955.62146 &16.456&15.656&15.206&16.336&16.266&15.806&16.176&16.026&15.506\\
 47955.65479 &16.396&15.436&15.146&16.406&16.346&15.946&16.266&15.896&15.696\\
 47955.67354 &16.436&15.426&15.276&16.346&16.336&16.066&16.256&15.816&15.746\\
 47956.40351 &15.906&15.596&16.526&      &      &15.936&15.536&16.086&16.236\\
 47956.42434 &16.016&15.656&16.496&      &      &15.806&15.046&16.116&16.246\\
 47956.44448 &16.106&15.716&16.396&      &      &15.606&15.136&16.146&16.206\\
 47956.48198 &16.236&15.826&16.486&      &      &15.426&15.396&16.156&16.276\\
 47956.50073 &16.296&15.916&      &      &      &15.396&15.526&16.146&16.326\\
 47956.51739 &16.346&15.946&16.466&      &      &15.516&15.616&16.146&16.246\\
 47956.53406 &16.406&15.996&16.606&      &      &15.656&15.726&16.156&16.246\\
 47956.55003 &16.436&16.056&16.526&      &      &15.776&15.796&16.126&16.286\\
 47956.56601 &16.456&16.076&16.056&      &      &15.856&15.856&16.096&16.306\\
 47956.58198 &16.476&16.096&15.306&      &      &15.956&15.946&16.046&16.366\\
 47956.59864 &16.456&16.076&14.966&      &      &15.956&16.016&15.996&16.376\\
 47956.61392 &16.456&16.106&15.056&      &      &      &16.066&15.936&16.376\\
 47956.62989 &16.436&16.106&15.216&      &      &16.126&16.106&15.876&16.256\\
 47956.64587 &16.446&16.146&15.436&      &      &16.316&16.176&15.836&15.876\\
 47956.66184 &16.436&16.156&15.496&      &      &16.256&16.216&15.796&15.266\\
 47956.67781 &16.396&16.176&15.626&      &      &16.296&16.266&15.766&14.846\\
 47960.48853 &15.806&15.934&15.461&16.083&16.353&16.275&15.856&16.112&16.166\\
 47960.50559 &15.891&15.976&15.611&15.751&16.255&16.286&15.301&16.066&16.218\\
 47962.42857 &15.641&15.771&15.661&      &      &      &16.213&      &16.242\\
 47962.44967 &15.191&15.854&15.818&      &      &      &16.231&      &16.232\\
 47962.47039 &15.261&15.908&15.933&      &      &      &16.222&      &16.361\\
 47962.49192 &15.431&15.958&16.077&      &      &      &16.343&      &16.380\\
 47962.51384 &15.601&16.004&16.190&      &      &      &16.207&      &16.258\\
 47962.53661 &15.751&16.055&16.275&      &      &      &15.852&      &16.070\\
 47962.59838 &16.046&16.123&16.535&      &      &      &15.241&      &14.951\\
 47962.62397 &16.137&16.173&16.566&      &      &      &15.411&      &15.131\\
 47962.64918 &16.239&16.236&16.623&      &      &      &15.601&      &15.301\\
 47962.67295 &16.347&16.232&16.580&      &      &      &15.751&      &15.431\\
 48659.61277 &16.380&      &16.389&16.098&16.353&16.068&15.631&16.132&15.561\\
 48659.63119 &16.435&      &16.423&16.101&16.345&16.126&15.701&16.152&15.651\\
 48659.64993 &16.468&      &16.504&16.182&16.397&16.159&15.787&16.179&15.751\\
 48659.70478 &16.457&16.049&16.545&      &16.430&16.264&15.969&16.229&15.973\\
 49803.43526 &16.246&16.089&15.441&16.310&16.037&16.227&      &      &16.244\\
 49803.50956 &16.441&16.183&15.973&16.363&15.341&      &      &      &16.292\\
 49803.67554 &16.657&      &      &      &      &16.308&      &      &      \\
 49809.44081 &15.391&16.119&16.522&      &16.438&15.701&      &      &16.260\\
 49810.40956 &15.982&15.701&16.471&16.344&16.318&15.451&      &      &16.000\\
\hline
\end{tabular}
\label{tab-varmag}
\end{center}
\end{table*}

\section{Periods and Light Curves}

\subsection{Periods}

The RR Lyrae variables in our target list are all the variables listed 
by SH73 lying 
in the field of view covered by the Loiano equipment, that were sufficiently 
un-crowded to allow photometric measurements of acceptable quality. 
Most of these stars 
had previous photometry, and a very accurate and comprehensive study of 
period variations was made by Szeidl (1965, 1973, hereafter S65 and S73 
respectively). 
Nemec and Clement (1989, hereafter NC89) 
analysed a number of RRc stars in a systematic search for double-mode 
pulsators. With the help of these results and the present photometry 
we have reanalysed the periods for all the program stars, using 
a period-finding program based on Stellingwerf's (1978) method  and 
hand computations based on the phase shifts between 
the various epochs of our observations. In this final step the time range 
spanned by our observations (i.e. 5 years) was very useful. 

We did not perform a detailed study of the double-mode pulsation 
characteristics for the two RRd variables V68 and V87. For these 
stars we simply adopted the period found by NC89 (V87) or a slightly 
improved value (V68). 

\subsection{Light-Curve Parameters} 

The B, V and I observations were fitted by smooth (``normal'') 
light curves in order 
to derive their parameters, i.e. mean magnitudes and colours, amplitudes 
and rise-times.  
The smoothing was performed using both Fourier-fitting techniques 
and a 3-rd order spline fitting to the data. 
However, because of the relatively small number of data 
points, these fitting techniques may give spurious results on 
those parts of the light curves where rapid 
changes occur. Therefore  the fits were inspected visually and the best 
eye fit was adopted on those parts of the light curve (typically around 
minimum and maximum light) where the smoothing techniques would fail. 

Mean B, V and I magnitudes were derived by averaging over the pulsation 
cycle the normal light 
curves both in intensity (then converted back to magnitudes) and 
in magnitudes directly. Mean magnitudes at minimum light were obtained 
by averaging in magnitude the normal light curves over the phase interval 
$0.5 \le \phi \le 0.8$. 
Amplitudes and rise-times (i.e. fraction of pulsation cycle from 
minimum to maximum light, $\Delta\phi$) were also obtained from the 
normal light curves. 

The adopted ephemerides and mean magnitudes are listed in 
Table~\ref{tab-efemer1}, and the amplitudes and rise-times 
are listed in  Table~\ref{tab-efemer2}. 

\begin{table*}
\caption{ Coordinates, periods, epochs and mean magnitudes of RR Lyrae 
variables in M 3}
\begin{center}
\begin{tabular}{rlrrllcccccc}
\hline\hline
Var. &type&  X  &  Y  &  Period  & Epoch of Max. & $<B>$ & $<V>$ & $<I>$ & 
$<B>$ & $<V>$ & $<I>$ \\
 & & (arcsec) & (arcsec) & (days) & 2400000+ & $mag$ & $mag$ & $mag$ & 
$int$ & $int$ & $int$ \\ 
\hline
  6  &ab & -123.9 &  60.1 &0.5143298 &47962.4497 &16.113 &15.753 &15.367
&16.030 &15.701 &15.348\\
 10  &ab &  153.6 & 138.0 &0.5695412 &47958.6366 &16.079 &15.697 &15.189
&16.029 &15.668 &15.178\\
 24  &ab & -147.6 &  10.4 &0.6633755 &47955.6624 &15.980 &15.553 &15.105
&15.939 &15.530 &15.096\\
 25  &ab & -124.4 & -31.4 &0.4800608 &47956.5986 &16.112 &15.755 &15.381
&15.992 &15.684 &15.357\\
 27  &ab & -110.2 &-102.8 &0.5790561 &47960.5379 &16.136 &15.677 &15.243
&16.091 &15.650 &15.229\\
 28  &ab &  -25.0 &-105.8 &0.4706131 &47954.4548 &15.986 &15.705 &15.375
&15.895 &15.652 &15.350\\
 29  &c  &  -65.2 & -73.6 &0.3156772 &47956.4889 &15.864 &15.582 &15.256
&15.824 &15.559 &15.248\\
 30  &ab &  -36.5 &  58.0 &0.5120984 &47956.4243 &15.939 &15.648 &15.219
&15.865 &15.600 &15.185\\
 31  &ab &   33.1 &  65.1 &0.5807361 &48754.4358 &16.083 &15.677 &15.248
&15.990 &15.619 &15.223\\
 32  &ab &   11.8 &  60.1 &0.4953518 &48754.5274 &16.085 &15.681 &15.241
&15.992 &15.627 &15.221\\
 34  &ab &  135.4 & 170.2 &0.5591012 &48754.5435 &16.080 &15.716 &15.227
&16.004 &15.669 &15.207\\
 41  &c  &  -93.3 &  54.0 &0.4883426 &47956.7264 & & & & & & \\
 42  &ab &  -78.6 &  41.0 &0.5900960 &47955.5027 &15.871 &15.565 &15.179
&15.751 &15.497 &15.156\\
 43  &ab &   99.9 &  24.7 &0.5404673 &47958.5741 &16.125 &15.748 &15.299
&16.004 &15.672 &15.265\\
 46  &ab & -128.1 & -51.5 &0.6133827 &47955.4291 &16.150 &15.695 &15.233
&16.123 &15.681 &15.228\\
 47  &ab & -117.5 & -73.2 &0.5409128 &47956.4445 &16.031 &15.630 &15.243
&16.000 &15.617 &15.238\\
 57  &ab &  155.2 &  -0.2 &0.5121902 &47958.5216 &16.138 &15.780 &15.360
&16.057 &15.726 &15.339\\
 58  &ab &  -86.2 &  46.2 &0.5170539 &47955.5909 &15.948 &15.660 &15.262
&15.835 &15.596 &15.239\\
 66  &ab & -101.4 & 121.4 &0.6201556 &47954.4715 &16.010 &15.682 &15.170
&15.979 &15.665 &15.164\\
 67  &ab & -131.4 & 123.0 &0.5683367 &47954.6076 &16.091 &15.726 &15.261
&16.029 &15.692 &15.248\\
 68  &d  &   21.9 & 174.8 &0.355634  &48750.3867 &16.070 &15.716 &15.292
&16.059 &15.706 &15.288\\
 69  &ab &   80.6 & 141.0 &0.5666159 &48746.4162 &16.140 &15.763 &15.278
&16.087 &15.730 &15.264\\
 70  &c  &   37.6 & 152.2 &0.4865540 &48746.4076 &15.739 &15.426 &14.998
&15.727 &15.420 &14.996\\
 74  &ab &   88.2 & 151.0 &0.4921490 &48754.6420 &(16.180) &(15.818) &(15.384)
&(16.102) &(15.762) &(15.365) \\
 75  &c  &   49.0 & 159.5 &0.3140800 &48750.3770 &15.966 &15.702 &15.363
&15.943 &15.687 &15.357\\
 76  &ab &  -14.4 & -88.2 &0.5017616 &47955.5833 &16.070 &15.742 &15.387
&15.952 &15.673 &15.363\\
 77  &ab &  -94.4 &  27.8 &0.4593505 &47956.4320 &16.097 &15.796 &15.455
&15.977 &15.727 &15.428\\
 78  &ab &   47.5 &  66.4 &0.6119254 &48750.5169 &16.074 &15.642 &15.182
&16.016 &15.606 &15.169\\
 84  &ab &   64.0 & 165.2 &0.5957289 &47958.4898 &16.108 &15.718 &15.203
&16.072 &15.699 &15.195\\
 87  &d  &  110.6 &  60.2 &0.357478  &48750.5169 &15.929 &15.622 &15.243
&15.908 &15.611 &15.238\\
 88  &c  &  -35.0 & -70.2 &0.2989933 &47954.4548 &15.825 &15.500 &(15.145)
&15.799 &15.488  & (15.141) \\
100  &ab &   69.9 &  97.3 &0.6188126 &48754.4878 &16.154 &15.746 &15.249
&16.126 &15.729 &15.242\\
101  &ab &   46.4 &  83.7 &0.6438975 &48746.3799 &16.221 &15.778 &15.278
&16.205 &15.768 &15.273\\
109  &ab &  -89.3 &   2.7 &0.5339175 &47956.6049 &16.087 &15.735 &15.312
&15.994 &15.682 &15.293\\
110  &ab &  -99.4 & -15.8 &0.5354569 &47956.4445 &16.049 &15.672 &15.270
&15.970 &15.630 &15.255\\
111  &ab &  -92.7 &  21.9 &0.5101896 &47954.4645 &16.039 &15.727 &15.309
&15.975 &15.690 &15.297\\
121  &ab &  -43.6 &  56.1 &0.5352090 &47954.7079 &16.081 &15.776 &15.334
&15.996 &15.729 &15.313\\
128  &c  &  114.6 & 131.4 &0.2920420 &47958.6575 &15.940 &15.696 &15.380
&15.912 &15.678 &15.374\\
129  &c  &  -43.6 &  72.2 &0.4112903 &47954.6277 &15.764 &15.493 &15.052
&15.747 &15.484 &15.049\\
130  &ab &    4.2 &  84.6 &0.5692737 &47956.4243 &15.949 &15.594 &15.168
&15.914 &15.574 &15.160\\
131  &c  &  -73.2 &  27.4 &0.2976886 &47955.6027 &15.874 &15.656 &15.428
&15.848 &15.640 &15.421\\
132  &c  &  -53.6 & -22.0 &0.3398479 &47955.6735 &15.959 &15.664 &15.269
&15.939 &15.653 &15.265\\
133  &ab &  -58.6 &  43.5 &0.5507230 &47954.6492 &16.103 &15.788 &15.116
&16.013 &15.742 &15.090\\
134  &ab &  -22.4 &  52.4 &0.6151824 &47956.5420 &16.040 &15.672 &15.284
&16.007 &15.654 &15.273  \\
135  &ab &  -27.0 &  38.0 &0.5683966 &47956.6521 &15.953 &15.578 &15.196
&15.897 &15.551 &15.183\\
136  &ab &  -25.4 &  33.4 &0.6175889 &47956.3999 &15.870 &15.448 &14.923
&15.844 &15.435 &14.917\\
140  &c  &  -15.7 & 108.9 &0.3331304 &47955.4048 &15.706 &15.491 &15.186
&15.686 &15.479 &15.181\\
142  &ab &  -30.0 & -59.0 &0.5686256 &47956.6299 &16.136 &15.784 &15.240
&16.060 &15.741 &15.227 \\
143  &ab &  -34.0 &  16.0 &0.5913771 &47955.4645 &15.879 &15.490 &15.150
&15.760 &15.428 &15.123\\
146  &ab &   96.0 & -59.0 &0.596758  &48750.4846 &16.119 &15.647 &15.118
&16.061 &15.618 &15.110 \\
149  &ab &   34.0 &  52.0 &0.5496744 &48751.4956 &16.136 &15.768 &15.355
&16.077 &15.729 &15.336 \\
150  &ab &   69.0 &  37.0 &0.5239411 &48751.4550 &16.188 &15.784 &15.350
&16.123 &15.740 &15.331\\
155  &c  &  -64.0 & -74.0 &0.3316733 &47955.6735 &15.807 &15.564 &15.225
&15.785 &15.550 &15.217 \\
167  &ab &  -97.0 &  -8.0 &0.6439944 &47955.6027 &16.099 &15.667 &15.178
&16.082 &15.658 &15.174 \\
168  &c  &  -78.0 & -37.0 &0.2764736 &47956.5643 &15.794 &15.592 &15.402
&15.781 &15.585 &15.398 \\
170  &c  &  -29.0 & -35.0 &0.4318107 &47955.5909 &15.548 &15.279 &14.944
&15.530 &15.270 &14.940\\
178  &c  &   79.0 &  46.0 &0.2670435 &48750.4673 &16.034 &15.799 &15.599
&16.019 &15.791 &15.594\\
188  &c  &  -15.0 &  32.0 &0.2662614 &47955.4833 &15.942 &15.768 &(15.589)
&15.932 &15.760 &(15.584)\\
197  &ab &   58.0 &  10.0 &0.4999345 &47958.4143 & & & & & &\\
209  &c  &  -13.0 & -29.0 &0.3484460 &47955.6027 &15.250 &15.071 &      
&15.242 &15.066 & \\
\hline
\end{tabular}
\label{tab-efemer1}
\end{center}
\end{table*}

\begin{table*}
\caption{ Mean magnitudes at minimum light, amplitudes and rise-times 
of RR Lyrae variables in M 3}
\begin{center}
\begin{tabular}{rlccccccccc}
\hline\hline
Var. &type& $B_{min}$ & $V_{min}$& $I_{min}$& A(B) &  A(V) &  A(I) & 
rt(B) & rt(V) & rt(I)\\
\hline
   6  &ab &16.436 &16.012 &15.515 &1.431 &1.141 &0.755 &0.11 &0.12 &0.12\\
  10  &ab &16.345 &15.903 &15.311 &1.072 &0.796 &0.528 &0.13 &0.13 &0.13\\
  24  &ab &16.249 &15.759 &15.232 &0.940 &0.737 &0.461 &0.19 &0.20 &0.20\\
  25  &ab &16.479 &16.042 &15.549 &1.647 &1.293 &0.841 &0.13 &0.13 &0.13\\
  27  &ab &16.391 &15.871 &15.374 &1.044 &0.853 &0.601 &0.13 &0.15 &0.15\\
  28  &ab &16.385 &16.010 &15.586 &1.277 &1.029 &0.693 &0.20 &0.20 &0.20\\
  29  &c  &16.162 &15.814 &15.390 &0.829 &0.653 &0.409 &0.40 &0.40 &0.40\\
  30  &ab &16.244 &15.907 &15.434 &1.310 &1.053 &0.966 &0.15 &0.15 &0.15\\
  31  &ab &16.411 &15.945 &15.435 &1.543 &1.204 &0.775 &0.14 &0.14 &0.14\\
  32  &ab &16.444 &15.954 &15.400 &1.408 &1.119 &0.724 &0.11 &0.11 &0.11\\
  34  &ab &16.416 &15.991 &15.414 &1.224 &1.003 &0.679 &0.18 &0.18 &0.18\\
  41  &c  &16.086 &15.792 &15.437 & & & & & &\\
  42  &ab &16.253 &15.854 &15.379 &1.595 &1.244 &0.735 &0.13 &0.13 &0.15\\
  43  &ab &16.510 &16.062 &15.505 &1.709 &1.396 &0.966 &0.12 &0.12 &0.13\\
  46  &ab &16.382 &15.867 &15.341 &0.793 &0.546 &0.380 &0.19 &0.20 &0.20\\
  47  &ab &16.306 &15.814 &15.363 &0.828 &0.549 &0.336 &0.32 &0.32 &0.29\\
  57  &ab &16.461 &16.045 &15.516 &1.375 &1.136 &0.786 &0.12 &0.11 &0.10\\
  58  &ab &16.329 &15.940 &15.435 &1.569 &1.235 &0.805 &0.14 &0.14 &0.14\\
  66  &ab &16.265 &15.868 &15.304 &0.789 &0.615 &0.362 &0.20 &0.23 &0.30\\
  67  &ab &16.386 &15.937 &15.404 &1.173 &0.948 &0.591 &0.16 &0.16 &0.16\\
  68  &d  &16.218 &15.860 &15.391 &0.462 &0.434 &0.263 &0.45 &0.45 &0.45\\
  69  &ab &16.397 &15.969 &15.413 &1.173 &0.938 &0.601 &0.15 &0.15 &0.15\\
  70  &c  &15.841 &15.508 &15.060 &0.480 &0.356 &0.237 &0.50 &0.50 &0.50\\
  74  &ab &16.484 &16.071 &15.546 &(1.294) &(1.190) &(0.715) &(0.10) &(0.11)
&(0.13)\\
  75  &c  &16.194 &15.899 &15.505 &0.617 &0.498 &0.323 &0.43 &0.41 &0.39\\
  76  &ab &16.467 &16.040 &15.572 &1.605 &1.287 &0.774 &0.15 &0.15 &0.15\\
  77  &ab &16.492 &16.108 &15.650 &1.581 &1.224 &0.807 &0.12 &0.14 &0.12\\
  78  &ab &16.335 &15.862 &15.308 &1.167 &0.946 &0.626 &0.12 &0.12 &0.12\\
  84  &ab &16.337 &15.892 &15.315 &0.939 &0.697 &0.438 &0.15 &0.15 &0.15\\
  87  &d  &16.171 &15.802 &15.361 &0.619 &0.463 &0.292 &0.25 &0.25 &0.25\\
  88  &c  &16.076 &15.674 &(15.239) &0.640 &0.457 &(0.254) &0.40 &0.40
&(0.40)\\
 100  &ab &16.387 &15.930 &15.369 &0.800 &0.642 &0.439 &0.18 &0.19 &0.18\\
 101  &ab &16.428 &15.934 &15.387 &0.589 &0.439 &0.345 &0.23 &0.23 &0.23\\
 109  &ab &16.410 &15.985 &15.466 &1.494 &1.156 &0.727 &0.11 &0.12 &0.13\\
 110  &ab &16.431 &15.941 &15.438 &1.309 &0.879 &0.550 &0.18 &0.18 &0.18\\
 111  &ab &(16.397) &(16.013) &(15.488) &1.115 &0.840 &0.483 &0.20 &0.20 &0.21\\
 121  &ab &16.483 &16.087 &15.558 &(1.288) &(0.927) &(0.626) &(0.20) &(0.20)
&(0.20)\\
 128  &c  &16.231 &15.929 &15.529 &0.709 &0.554 &0.331 &0.35 &0.35 &0.35\\
 129  &c  &15.822 &15.565 &15.115 &0.558 &0.427 &0.265 &0.52 &0.50 &0.50\\
 130  &ab &16.165 &15.758 &15.280 &0.893 &0.686 &0.461 &0.15 &0.15 &0.16\\
 131  &c  &16.130 &15.864 &15.562 &0.682 &0.546 &0.334 &0.30 &0.30 &0.30\\
 132  &c  &16.109 &15.797 &15.351 &0.574 &0.455 &0.240 &0.40 &0.40 &0.40\\
 133  &ab &16.449 &16.036 &15.326 &1.364 &1.051 &0.805 &0.13 &0.13 &0.15\\
 134  &ab &16.285 &15.867 &15.435 &0.876 &0.635 &0.526 &0.19 &0.19 &0.19\\
 135  &ab &16.195 &15.752 &15.329 &1.138 &0.902 &0.610 &0.15 &0.16 &0.15\\
 136  &ab &16.079 &15.601 &15.046 &0.780 &0.553 &0.388 &0.15 &0.16 &0.17\\
 140  &c  &15.863 &15.620 &15.265 &0.604 &0.468 &0.299 &0.42 &0.42 &0.42\\
 142  &ab &16.442 &16.011 &15.356 &1.417 &1.000 &0.630 &0.12 &0.13 &0.14\\
 143  &ab &16.290 &15.791 &15.321 &1.561 &1.204 &0.878 &0.12 &0.12 &0.12\\
 146  &ab &16.333 &15.831 &15.220 &1.046 &0.809 &0.472 &0.15 &0.15 &0.15\\
 149  &ab &16.372 &16.005 &15.511 &1.238 &0.974 &0.659 &0.12 &0.12 &0.12\\
 150  &ab &16.505 &16.049 &15.524 &1.155 &0.989 &0.699 &0.13 &0.14 &0.14\\
 155  &c  &16.041 &15.748 &15.373 &0.602 &0.481 &0.359 &0.36 &0.36 &0.36\\
 167  &ab &16.283 &15.796 &15.263 &0.599 &0.443 &0.258 &0.20 &0.20 &0.20\\
 168  &c  &15.948 &15.730 &15.507 &0.475 &0.370 &0.290 &0.40 &0.40 &0.40\\
 170  &c  &15.662 &15.376 &15.025 &0.595 &0.439 &0.278 &0.50 &0.50 &0.50\\
 178  &c  &16.185 &15.907 &15.665 &0.515 &0.376 &0.290 &0.45 &0.45 &0.45\\
 188  &c  &16.064 &15.866 &(15.696)&0.439&0.370&(0.306)&0.45&0.45&(0.45)\\
 197  &ab &       &       &       &      &      &      &     &     &   \\
 209  &c  &15.351 &15.178 &       &0.375 &0.301 &      &0.40 &0.40 &   \\
\hline
\end{tabular}
\label{tab-efemer2}
\end{center}
\end{table*}

Columns 1, 2, 3 and 4 of Table~\ref{tab-efemer1} contain the star 
identification, type and position in arcsec from the cluster center, taken 
from SH73. The new periods and epochs are listed in columns 5 and 6 (the 
epochs are the Heliocentric Julian Days of maximum light). 
Columns 7-9 list the B, V and I mean magnitudes averaged in magnitude, 
and col. 10-12 list the mean magnitudes averaged in intensity. 
In Table~\ref{tab-efemer2}, col. 3-5 list the mean magnitudes at minimum light
(i.e. averaged in magnitude over the phase 
interval $0.5 \le \phi \le 0.8$), col. 6-8 list the light curve 
amplitudes in the three colours, and col. 9-11 list the rise-times 
(i.e. fraction of pulsation cycle from minimum to maximum light). 

Typical {\it internal} errors of the mean photometric values are on average 
$\le 0.03 mag$ for V and B, and $\le 0.05 mag$ for I. 
Values in parenthesis are particularly uncertain. 

\begin{figure*}
\vspace{20cm}
\caption{$B$, $V$ and $I$ light curves for 6 of the variables observed}
\includegraphics{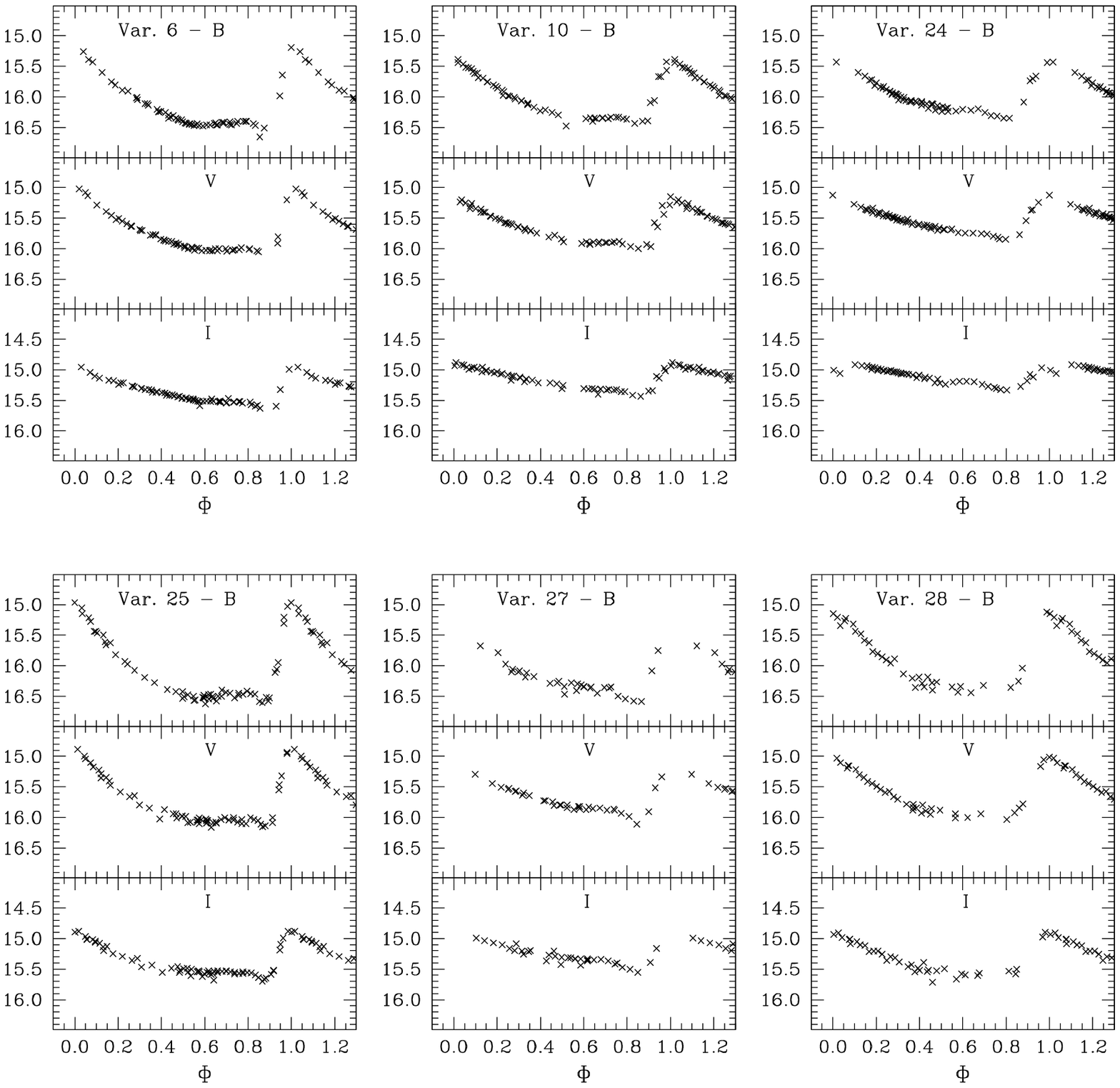}
\label{fig-f3}
\end{figure*}

Comments on the individual stars are presented in the following section. 
The B, V and I light curves are shown in Fig~\ref{fig-f3} through 
Fig~\ref{fig-f12}.

\begin{figure*}
\vspace{20cm}
\caption{see Figure 3}
\includegraphics{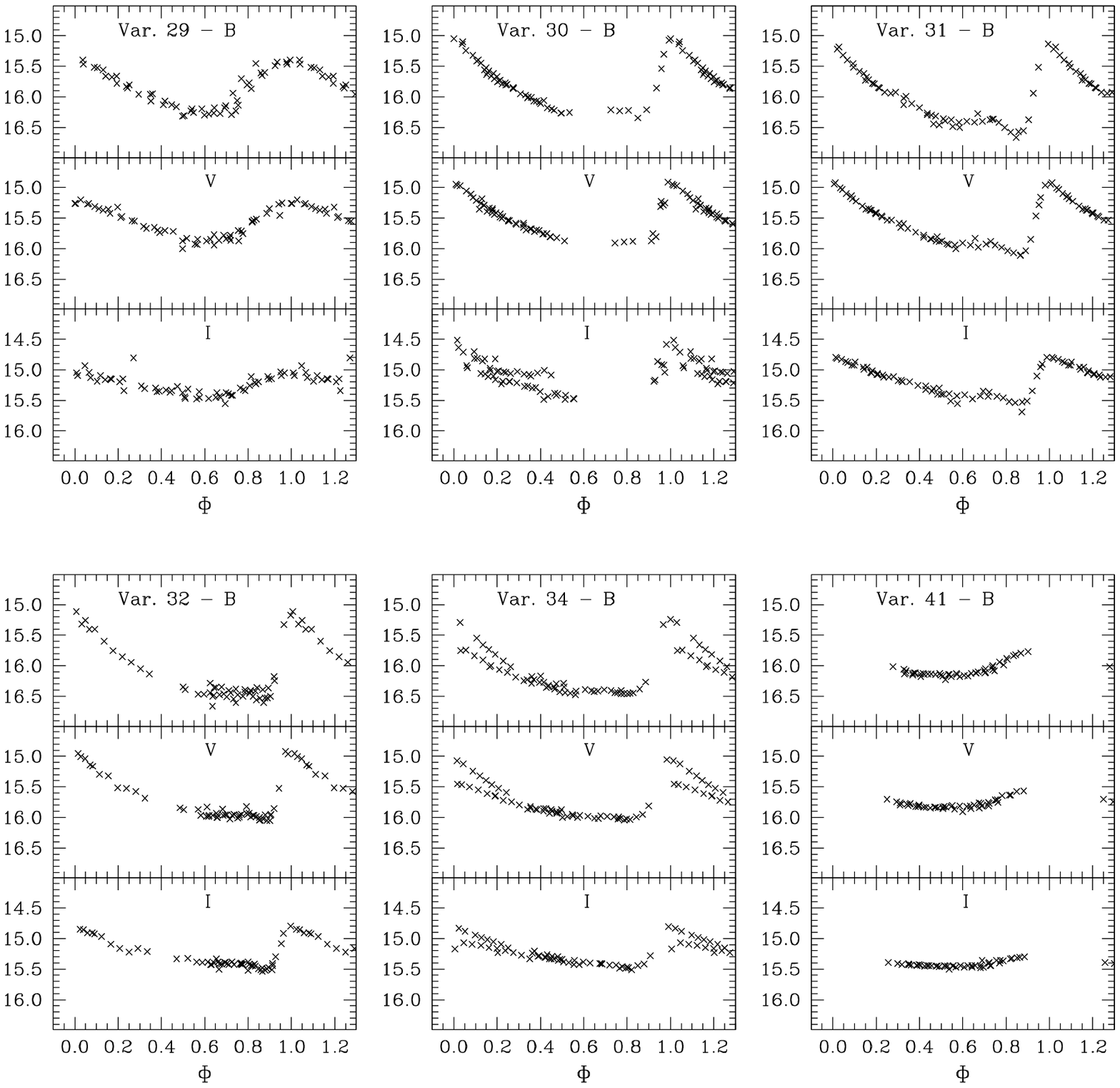}
\label{fig-f4}
\end{figure*}

\begin{figure*}
\vspace{20cm}
\caption{see Figure 3}
\includegraphics{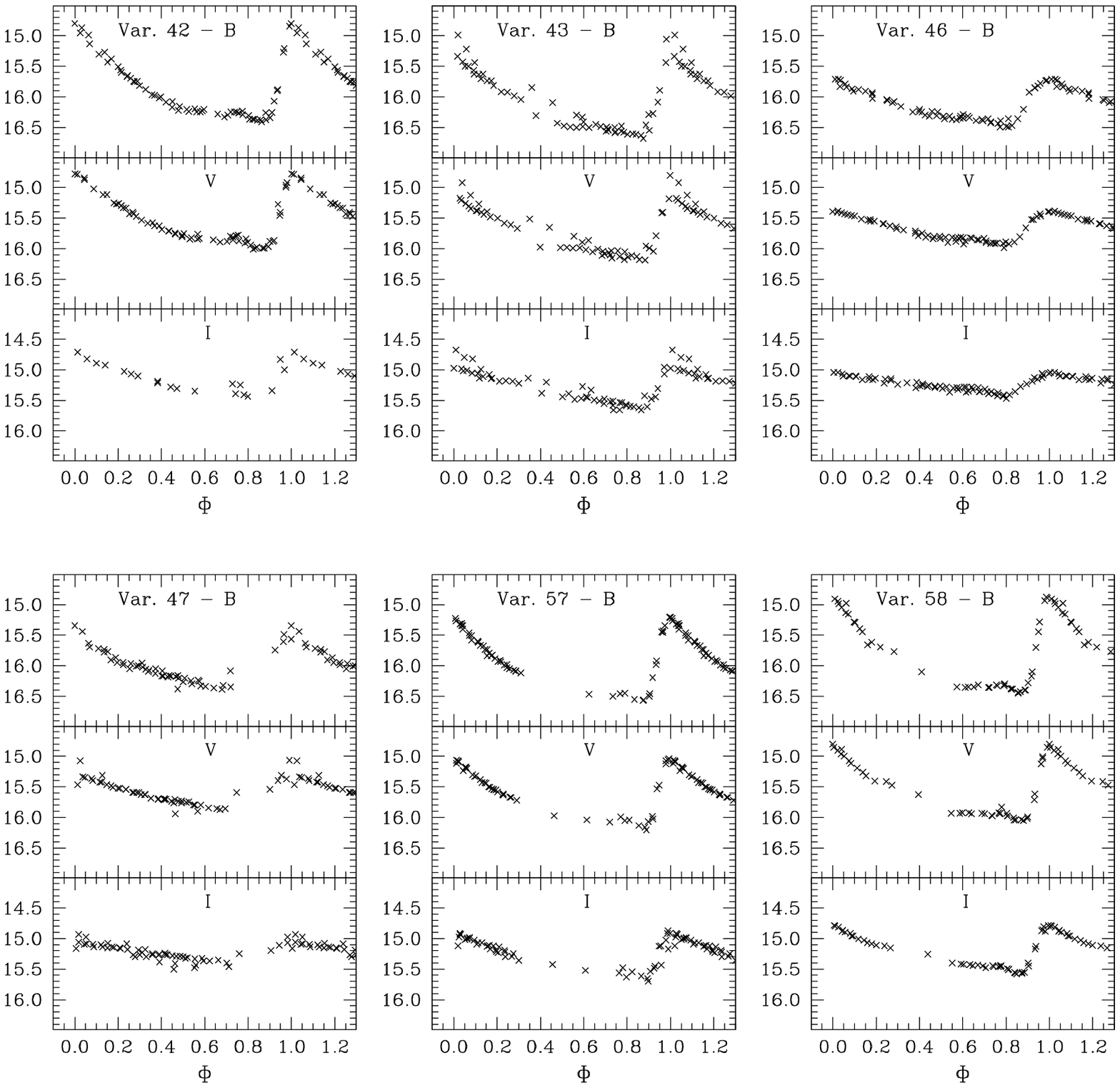}
\label{fig-f5}
\end{figure*}

\begin{figure*}
\vspace{20cm}
\caption{see Figure 3}
\includegraphics{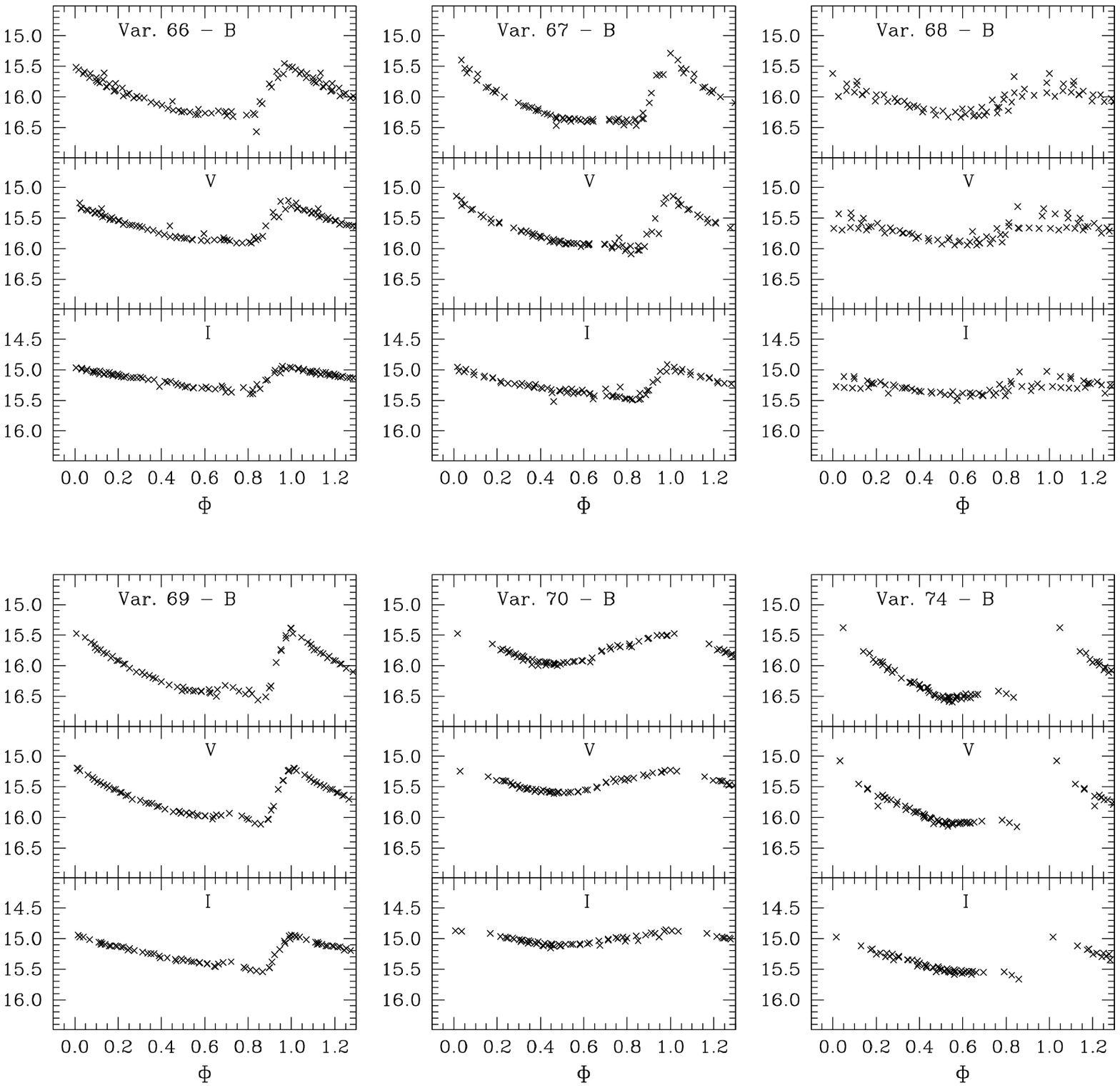}
\label{fig-f6}
\end{figure*}

\begin{figure*}
\vspace{20cm}
\caption{see Figure 3}
\includegraphics{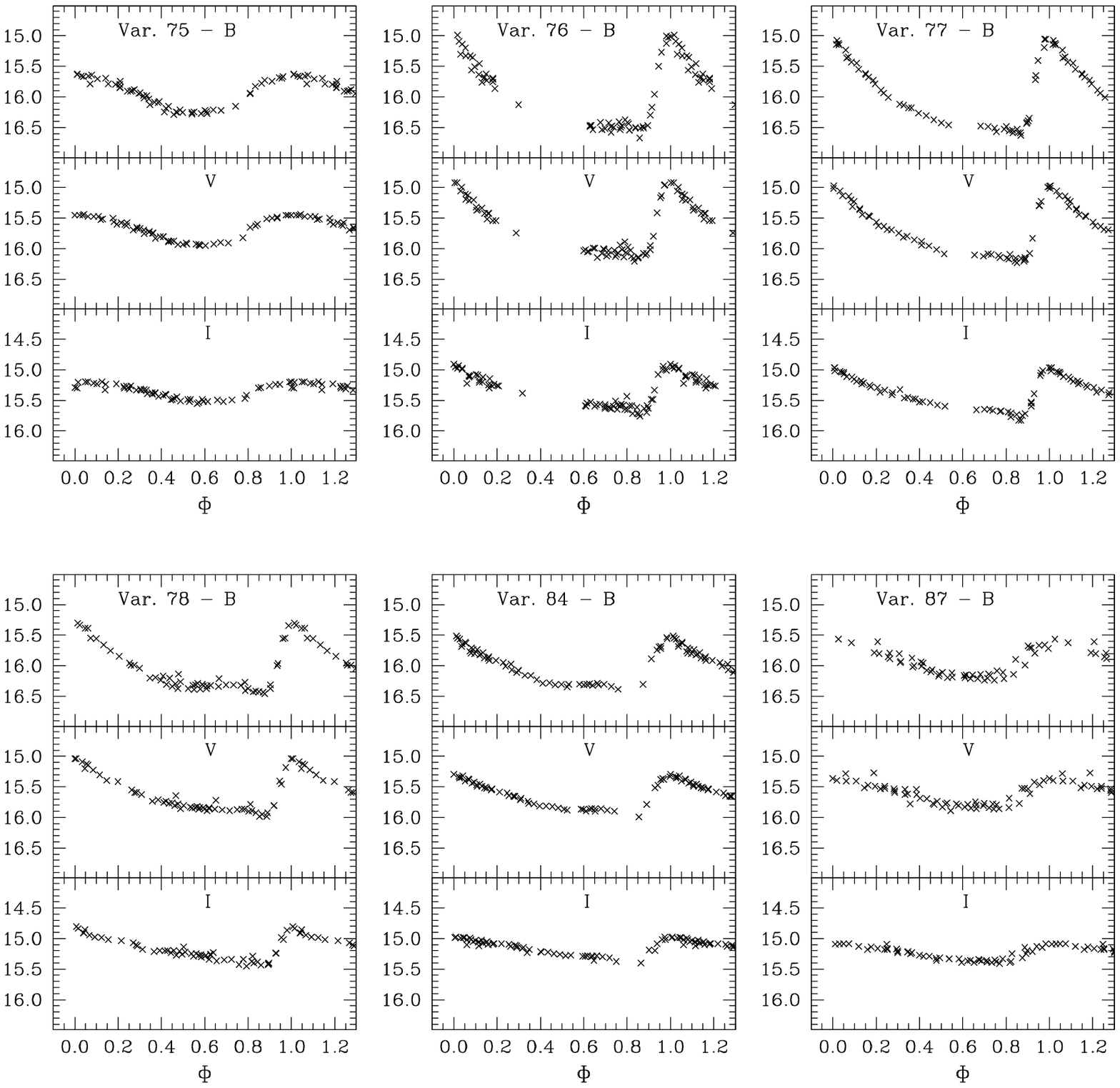}
\label{fig-f7}
\end{figure*}

\begin{figure*}
\vspace{20cm}
\caption{see Figure 3}
\includegraphics{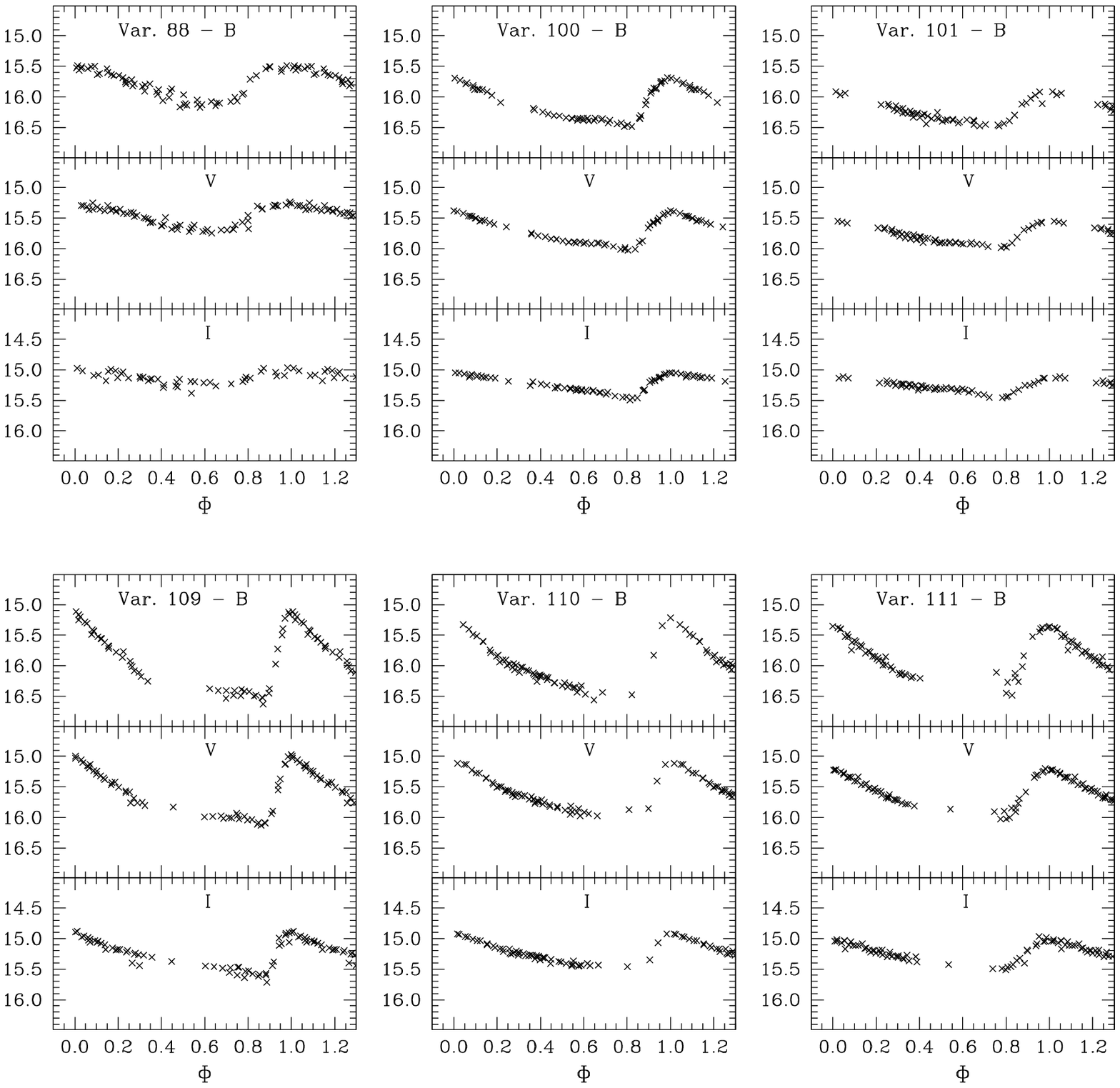}
\label{fig-f8}
\end{figure*}

\begin{figure*}
\vspace{20cm}
\caption{see Figure 3}
\includegraphics{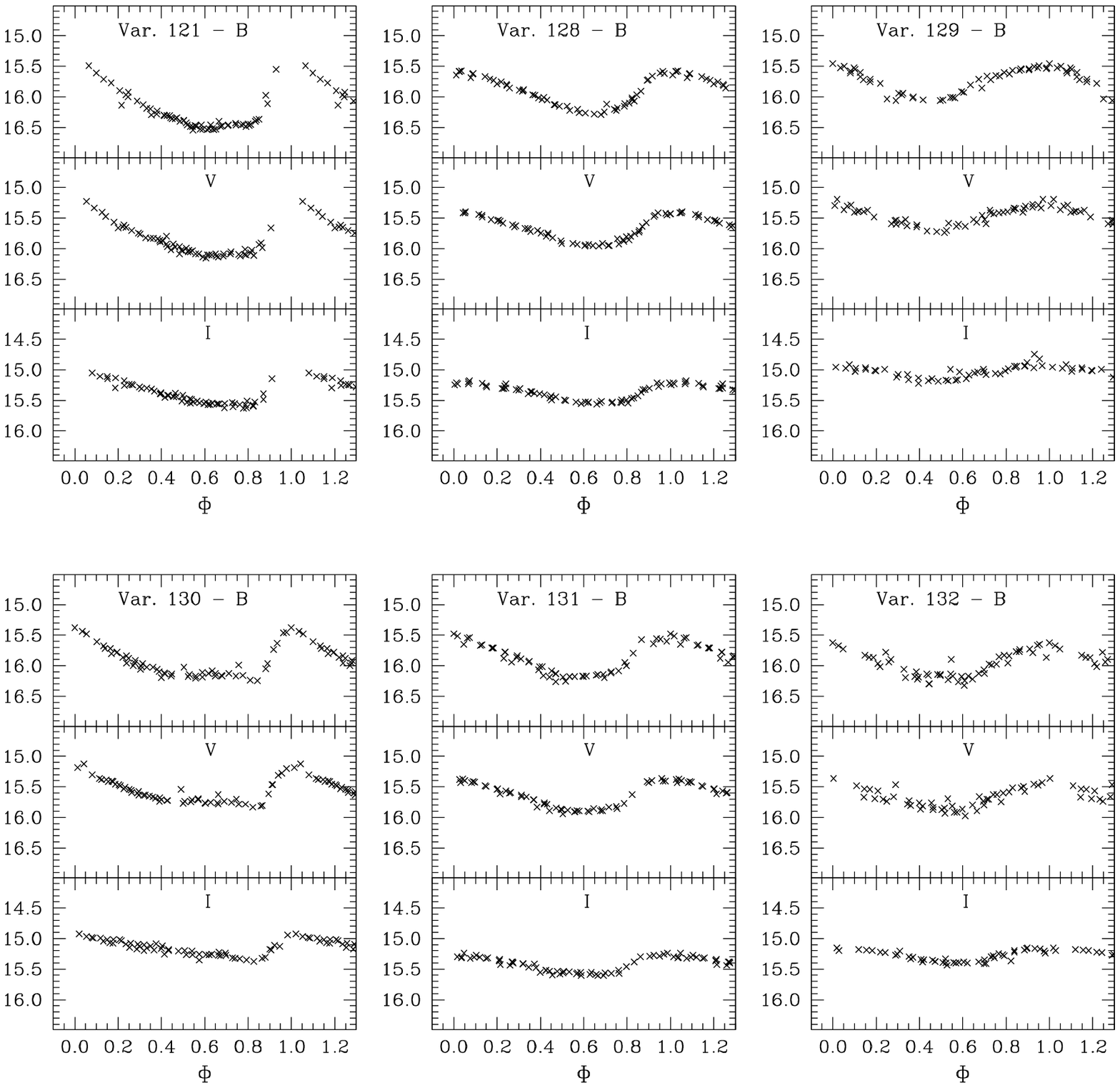}
\label{fig-f9}
\end{figure*}

\begin{figure*}
\vspace{20cm}
\caption{see Figure 3}
\includegraphics{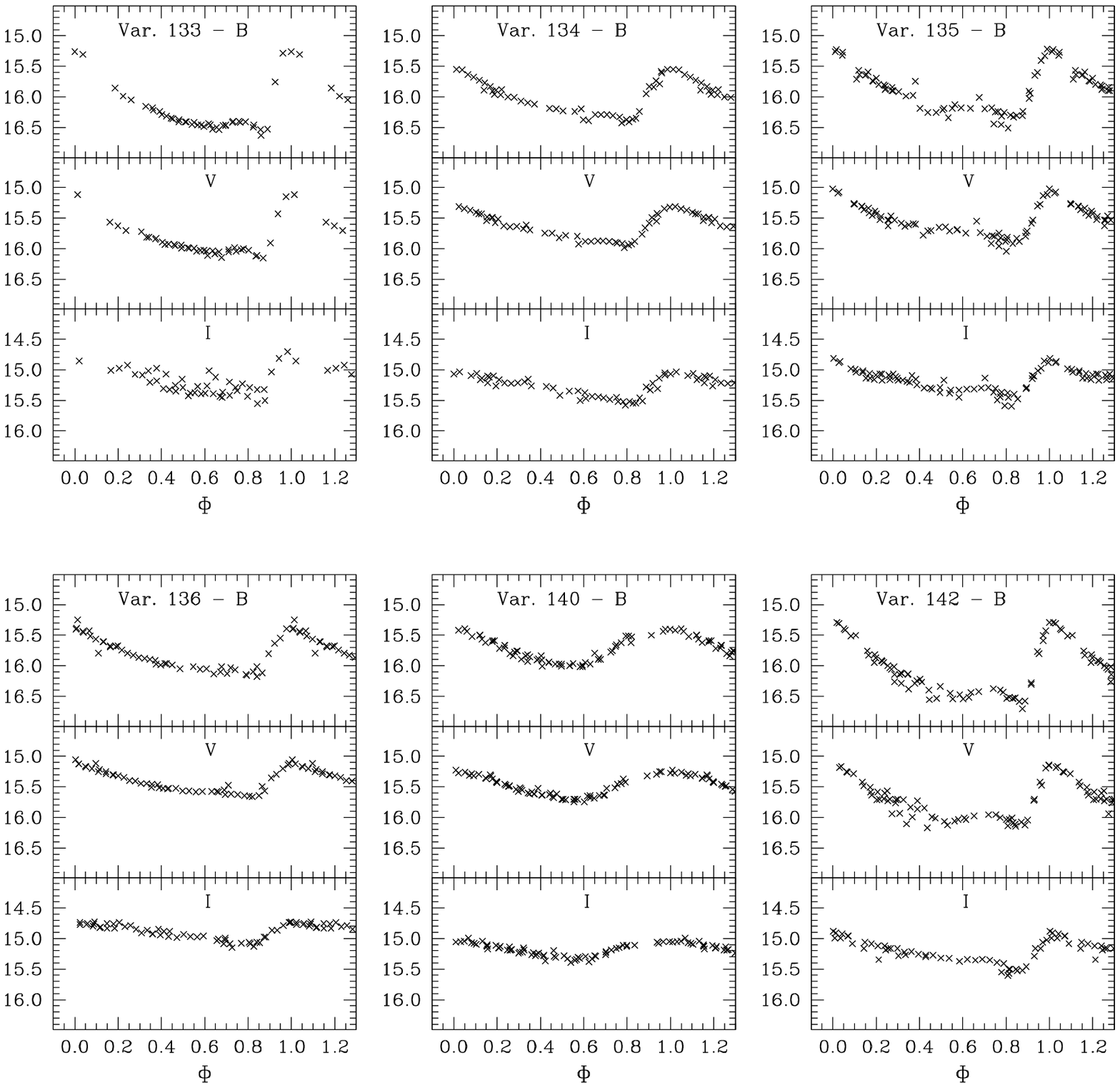}
\label{fig-f10}
\end{figure*}

\begin{figure*}
\vspace{20cm}
\caption{see Figure 3}
\includegraphics{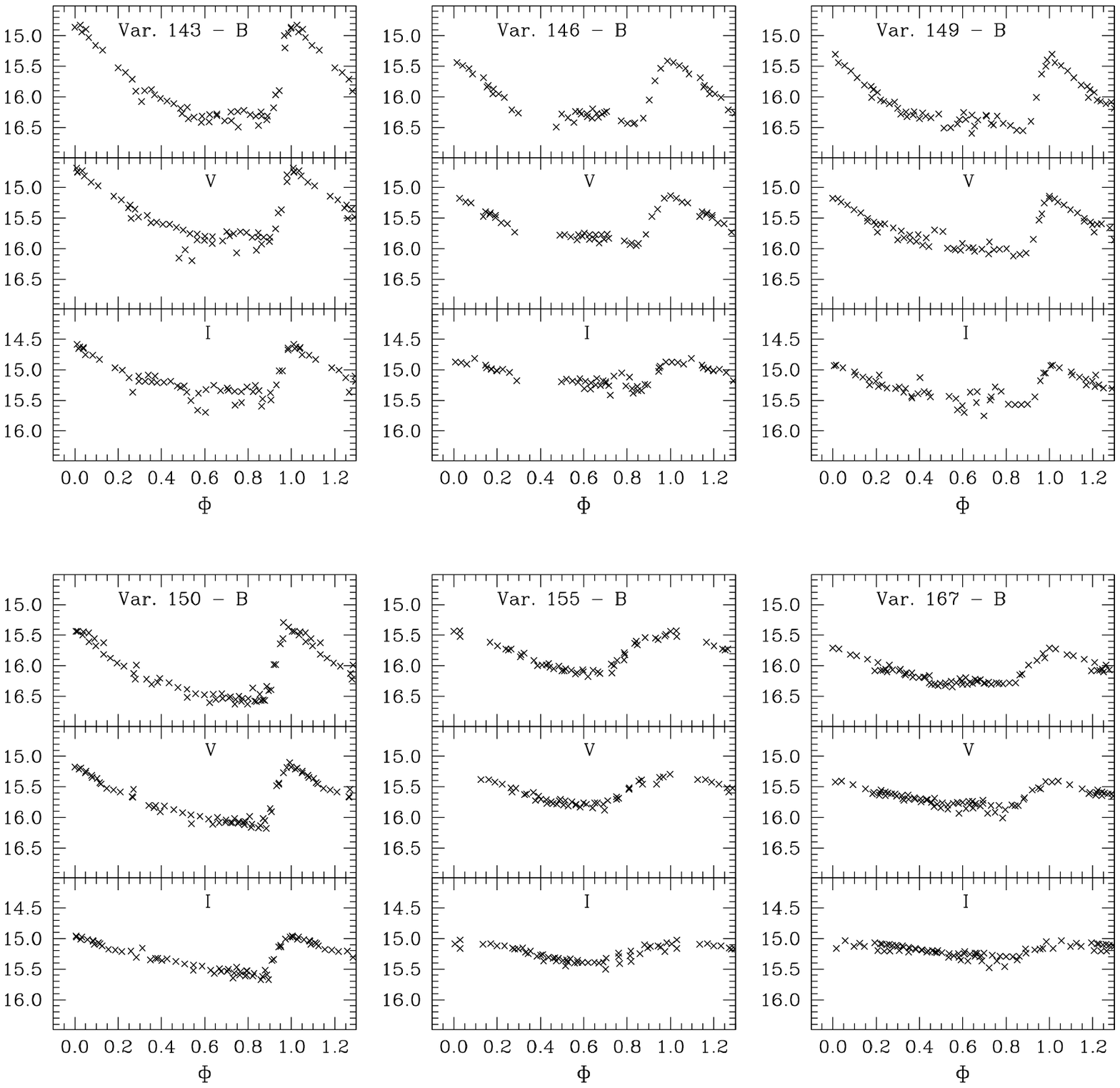}
\label{fig-f11}
\end{figure*}

\begin{figure*}
\vspace{20cm}
\caption{see Figure 3}
\includegraphics{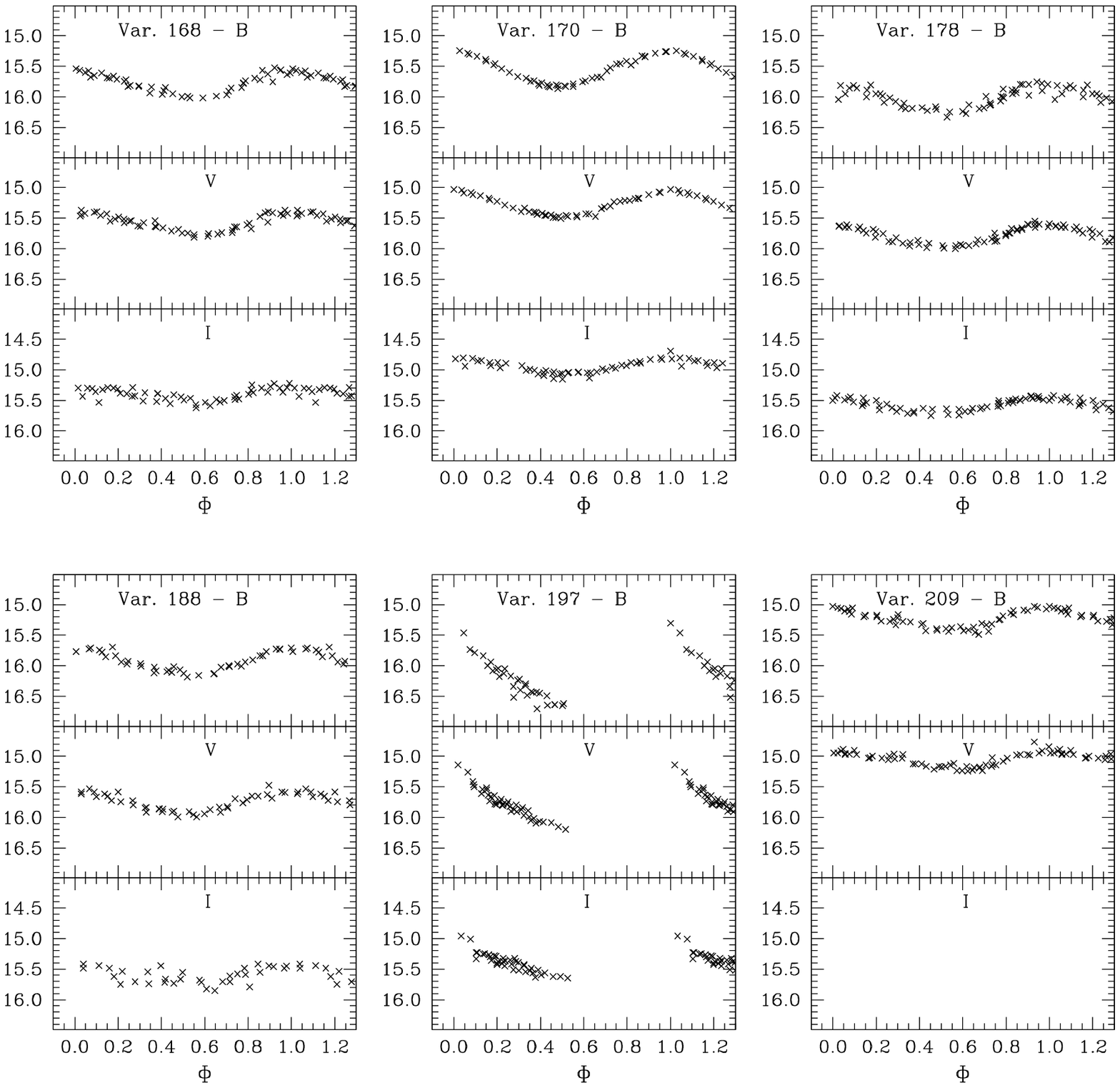}
\label{fig-f12}
\end{figure*}

\subsection{Notes on individual variables} 

{\bf V6}: The exact time of minimum light is missing in our observations. 
The minimum scatter in the light curves is obtained for a slightly longer 
period than used by S65, in agreement with the positive  O-C parabolic 
curve. 

{\bf V10}: The minimum scatter is obtained for a longer period than 
given by S65, in agreement with the positive O-C parabolic curve. 
There is no strong indication of the variations in the light curve shapes 
that were suggested by S65. 

{\bf V24}: Our best period is somewhat longer than S65, although his 
O-C was represented by a straight flat line (but ``smaller oscillations 
may be real''). Some scatter is present on the descending branch. 
Since the minimum 
and maximum light are (poorly) covered by data taken on one night only, 
it is not possible to assess the presence of Blazhko effect. 

{\bf V25}: Our best period is slightly longer than the period given by 
S65,  in agreement with the slightly positive O-C curve. 
There is some indication of larger than normal scatter on the descending 
branch and around minimum light. 

{\bf V27}: The period that best fits to our observations is shorter than 
that given by S65, in agreement with the negative O-C parabola. The 
maximum light is missing in our observations. 

{\bf V28}: The data available in the literature show a strongly varying 
light curve (S65). This star was suggested by NC89 as a possible candidate 
double-mode pulsator with primary period P=0.470120, but no secondary 
period could be found, so NC89 concluded that V28 is probably a Blazhko 
variable. The period that best fits to our observations is P=0.4706131, shorter 
than the value given by S65 and in agreement with the negative slope of the 
last oscillation in the O-C curve. Some scatter in the data, especially on the 
descending branch, does indeed suggest variations that could be due to 
Blazhko effect. 

{\bf V29}: this star has no previous period determination. 
It is very close to the center, and the quality of the photometry is 
quite poor. 

{\bf V30}: this variable was included in the study by S73. The best period
we have found is slightly longer than the value given by S73 and SH73,
in agreement with the O-C trend derived by S73.
The star is near the cluster center and has very close companions that 
disturb the photometry, especially in the I band where the effect of poor 
seeing is more severe. 

{\bf V31}: the period that best fits to our data is slightly longer than 
S65 period, in spite of a negative parabolic O-C (but an oscillating 
O-C was also possible). 

{\bf V32}: the period used by S65 is providing the best fit to our data as 
well, although the O-C was a slightly negative parabola. The star lies in 
a dense region near the center, and some of the scatter seen at minimum 
light may be due to larger photometric errors. 

{\bf V34}: the period used by S65 provides the best fit to our data as 
well. There is clear evidence of variability in the light curve shape
(Blazhko effect), 
especially at maximum light and on the descending branch, and the 
difference in the height of maxima amounts to about 0.4 mag, as 
already noticed by S65. The nights JD47958 and JD49803-10 seem to share 
the same Blazhko phase. The values in Tables 3 and 4 correspond to the 
``bright'' light curves. 

{\bf V41}: the period that best fits to our data is somewhat longer than the 
period used by S65, in agreement with the positive trend of the O-C 
curve. A fraction of the light curve is missing and no average parameters 
could be derived. 

{\bf V42}: our best period is smaller than the period used by S65, 
in agreement with the strongly negative O-C parabola. The light curves 
are well defined. 

{\bf V43}: we have found a period 0.5404673, somewhat shorter than the period 
used by S65, as expected from the negative O-C parabola. However this 
result is quite inaccurate since the 
light curves are strongly variable, with a $\Delta$V at maximum light of 
about 0.35 mag. The values in Tables 3 and 4 correspond to the 
``bright'' light curves.

{\bf V46}: our best period is slightly longer than the period given by 
S65, in agreement with the positive O-C curve. There is some scatter 
in the data, but no clear indication of light curve variability. 

{\bf V47}: the star has a variable light curve, as already commented by 
S65. Our best period puts in phase the maxima, which differ by about 
0.25 mag in V, but the Calar Alto data would suggest an even shorter 
period. 

{\bf V57}: our new period is shorter than the period given by S65, as 
expected from the strongly negative parabolic shape of the O-C curve. 
The scatter of the data is quite small and there is no indication of 
light curve variation. 

{\bf V58}: our new period is slightly shorter than the period adopted by S65, 
but the shape of the O-C diagram was irregular making any prediction
difficult. The scatter of the data is very small, the indication 
of light curve variation on the descending branch suggested by the 
Calar Alto data may not be significant. 

{\bf V66}: this variable has a varying light curve. 
Our best period is somewhat shorter than the period used by S65, 
but it is still quite uncertain. 

{\bf V67}: our best period is somewhat shorter than S65 adopted period, 
as expected from the negative parabolic shape of the O-C curve. 
The star shows indications of a variable light curve, although the 
variations in our data do not look as strong as noted by S65. 

{\bf V68}: double-mode pulsator. The primary period (0.47850) derived 
by NC89 does not fit well to our data. Our best-fit period (0.355634) 
is more similar to the secondary period found by NC89 (0.355978). 
The difference is significant. The overtone periods derived from previous 
data sets are 0.355978 for data taken in the 1920s (NC89), 0.355974 
for data taken in the 1940s (Martin 1942), and 0.355973 for data taken 
in the 1950s (Goranskij 1981). When compared to this study with P=0.355634 
for data taken in the 1990s, this seems to indicate a decrease. 

{\bf V69}: in agreement with the positive parabolic shape of the O-C 
curve, the period that best fits to our data is slightly longer than in S65. 
The variable could be resolved from its close companion in our photometry, 
and the light curves are very well defined with small scatter. 

{\bf V70}: the old data for this variable have been recently reanalysed 
by NC89, who derived a period 0.486085 and suggested this is a fundamental 
pulsator with unusually small amplitude. Our data do not provide a 
good light curve with NC89's period; we have derived a slightly longer 
period (0.486554) and the shape of the light curve is sinusoidal with 
a V amplitude about 0.35 mag, rather typical of a Bailey type-c variable. 
This is confirmed also by the K97 photometry. 

{\bf V74}: our best period is slightly longer than S65, in agreement with 
the mildly positive parabolic shape of the O-C curve. The exact times 
of minimum and maximum light have been missed in our observations, but 
the rest of the curve is very well defined with little scatter. Therefore 
the period is quite accurate, but the average magnitudes and amplitudes are 
not. 

{\bf V75}: this star is among those reanalysed by NC89. We find that the 
period derived by NC89 fits well also to our observations. 

{\bf V76}: our best period is slightly longer than the period used by S65, 
as expected from the mildly positive trend of the O-C curve. Some larger 
than normal scatter around minimum light and on the descending branch 
suggests possible variations of the light curve. 

{\bf V77}: our best period is slightly longer than the period used by S65, 
as expected from the mildly positive trend of the O-C curve. 

{\bf V78}: the period used by S65 seems to be fitting to our data as well, 
in spite of the mildly positive trend of the O-C curve. Some scatter 
along the descending branch (especially evident for the Calar Alto data) 
suggest light variations. 

{\bf V84}: the period used by S65 fits well also to our data. The scatter 
is quite small and the light curves are well defined with no indication 
of variations. 

{\bf V87}: double-mode pulsator. The period derived by NC89 fits well also 
to our data. 

{\bf V88}: this star was reanalysed by NC89, who derived a period 0.298985
based on old photometric data. The period that best fits to our data is 
somewhat longer, i.e. 0.2989933, compatible with the sinusoidal shape of 
the O-C curve (S65). The I data have a very large scatter, and the average 
I magnitudes and amplitude are therefore very uncertain. 

{\bf V100}: the period used by S65 fits very well also to our data. The scatter 
is small and the curves are very well defined. 

{\bf V101}: the period used by S65 fits very well also to our data. The scatter 
on the descending branch is somewhat larger than normal and may indicate 
light curve variations. 

{\bf V109}: our best period is slightly shorter than the period used by 
S65, and is compatible with the oscillating O-C curve. 

{\bf V110}: our best period is significantly longer than the period given by
S65, indicating that the trend of the O-C curve has reversed. 
The light curves are well defined and there is no significant indication of 
variations, however the minimum light has been missed by our observations. 

{\bf V111}: This star was studied by S73, who found strong variations in
the light curve and a significantly negative O-C. In agreement with that,
our best period is shorter than the value listed by S73 and SH73; we also
find indication of Blazhko effect.

{\bf V121}: our best period is slightly longer than the period used by 
S65, in agreement with the oscillating O-C curve. The exact time of 
maximum light is missing in our observations. 

{\bf V128}: This star was studied by S73 who found a period P=0.292271
days, and was later analysed by NC89 who found a period 0.292042 days.
This latter period fits to our data as well.

{\bf V129}: This star is not among those studied by S65 and S73, but was
analysed by NC89 who derived a period 0.305474.
They noted, however, that ``... the
available photometry only weakly constrains the period ... The ... Larink
and Greenstein's data cannot distinguish between 0.30d or 0.43d.''
With our data we can clearly rule out a period around 0.30d; the period
that best fits to our observations is 0.4112903.
   
{\bf V130}: This star was studied by S73, who derived a period P=0.5688172
days, noting however that ``... P=0.569665 seems almost just as good''.
The O-C was significantly negative. Our best period P=0.5692737 seems
to indicate that the alternative (longer) period found by S73 was probably
the correct one.  The indication that the
star may show the Blazhko effect is confirmed, although only marginally,
by our data.
           
{\bf V131}: This star has been studied in detail both by S65 and NC89, 
who suggest the period 0.2976919. Although only marginally different, 
the period that best fits to our data is 0.2976886. 

{\bf V132}: This star was not included in the studies by S65 and S73,
but was reanalysed by NC89 who derived the period 0.339825 days. However,
our data provide a somewhat better fit using the older period listed
by SH73, namely 0.3398479 days.
 
{\bf V133}: This star was not included in the studies by S65 and S73.
The period
listed by SH73 provides a good fit to our data as well, however our
photometry does not put much constraint on the determination of a new
period.

{\bf V134}: This star was not included in the studies by S65 and S73.
The period
could be slightly improved with respect to the period listed by SH73.
The scatter in the data may suggest light curve variations.

{\bf V135}: This star was not included in the studies by S65 and S73.
Our data
suggest a period certainly shorter than the period listed by SH73.
A rather large scatter in the photometry may suggest light curve
variations.
   
{\bf V136}: This star had no previous period determination as it lies 
very close to the cluster center. 

{\bf V140}: The star was analysed both by S65 and NC89, and their best 
period also fits well to our data. NC89 commented that ``... there seem to be 
some real irregularities in the light curves. For example, there is a 
phase shift between ...'' different sets of observations. 
The photometric scatter of our data does not allow us to confirm or 
reject this suggestion. 

{\bf V142}: The period used by S65 provides a good fit to our data as 
well. The scatter on the descending branch seems to indicate variations 
in the shape of the light curves.

{\bf V143}: This star was not included in the studies by S65 and S73. Our
best period is significantly longer than the period listed by SH73. The
star is quite close to the center, some of the large scatter around minimum
light may be due to photometric errors.

{\bf V146}: This star was not included in the studies by S65 and S73. The
period listed by SH73 (0.596740 days) fits to our data as well, however our
photometry for this star has a time baseline of 8 days only, therefore does
not constrain any new period determination. This star was later analysed by
Kholopov (1977), however his updated period (0.502193 days) does not fit to
our data.
 
{\bf V149}: This star was not included in the studies by S65 and S73.
Our best period (0.5496744 days) is significantly shorter than the period
listed
by SH73 (0.54985 days). The scatter on the descending branch is rather large,
especially in the I-band.

{\bf V150}: This star was not included in the studies by S65 and S73.
Our best period (0.5239411 days) is somewhat shorter than the period listed
by SH73 (0.52397 days).
  
{\bf V155}: This star had no previous period determination. 

{\bf V167}: This star was included in the study by S73.
The period that best fits to our data is slightly longer than the period
listed by S73 and SH73.
There is significant scatter along the descending branch.

{\bf V168}: This star was not included in the studies by S65 and S73. It was
reanalysed by NC89, and we confirm their conclusion that the period
is less than 0.30 d. We find the minimum scatter in the light curves
using a period somewhat longer than the period derived by NC89.
  
{\bf V170}: This star was not included in the study by S65 and S73. It was
reanalysed by NC89, who derived a period of 0.43771d, with possible
alias periods of 0.29702 and 0.6774 days. Based on our photometric data
we can exclude the alias periods; the period that best fits to our data
is 0.4318107 which is somewhat shorter than the period listed by SH73 
(i.e. 0.43725). The shape of the light curve is definitely sinusoidal,
suggesting that the star is a type-c variable with an unusually long
period.
 
{\bf V178}: This star was included in the study by S73, who derived
the period  P=0.2651549 days for the years 1925-1941, and P=0.2650805 days
for the years 1950-1962, with a sudden increase around 1956. It was
reanalysed by NC89, who derived a period of 0.26498 days. The period that
best fits to our data is 0.2670435 days.

{\bf V188}: This star was not included in the studies by S65 and S73. It was
reanalysed by NC89, who derived a period of 0.2706 days. The period that best
fits to our data is 0.2662614 days, however the time baseline covered by our
observations is only 3 days, and does not put much constraint on the
period determination. The scatter in the I data is quite large and the
average I magnitude and amplitude are therefore very uncertain.
   
{\bf V197}: This star was not included in the studies by S65 and S73.
The period that best fits to our data is somewhat shorter than the
period listed by SH73. The light curves are defined only for part of the
descending branch, and the scatter in the data is large; the minimum
light, ascending branch and maximum light are missing, therefore it is
impossible to derive reliable mean magnitudes and amplitudes.

{\bf V209}: This star has no previous period determination. 
No reliable I data could be obtained, possibly because of the central 
location and severe crowding.

\bigskip\bigskip\noindent
ACKNOWLEDGMENTS

\noindent
It is a pleasure to thank M. Corwin for several helpful discussions. We are 
very grateful to J. Kaluzny for allowing us to use their 
unpublished data to check our photometric calibrations.

\end{document}